\documentclass[
    ,final            % use final for the camera ready runs
  ] {aipproc}

\layoutstyle{6x9}
\usepackage{amssymb}
\begin{document}

\title{The ontology of General Relativity}

              

\author{Gustavo E. Romero}{address={Instituto Argentino de Radioastronomía (CCT- La Plata, CONICET), C.C.5, 1894 Villa Elisa, Buenos Aires, Argentina.},
  altaddress={Facultad de Ciencias Astronómicas y Geofísicas, Universidad Nacional de La Plata, Paseo del Bosque, 1900, La Plata, Argentina.}}

%\author{<author3>}{
%  address={<common address for author2 and author3>}
%  ,altaddress={<author1 address>} % additional visiting address
%}

\begin{abstract}
I discuss the ontological assumptions and implications of General Relativity.  I maintain that General Relativity is a theory about gravitational fields, not about space-time. The latter is a more basic ontological category, that emerges from physical relations among all existents. I also argue that there are no physical singularities in space-time. Singular space-time models do not belong to the ontology of the world: they are not things but concepts, i.e. defective solutions of Einstein's field equations. I briefly discuss the actual implication of the so-called singularity theorems in General Relativity and some problems related to ontological assumptions of Quantum Gravity.

\end{abstract}

\maketitle

%%%%%%%%%%%%%%%%%%%%%%%%%%%%%%%%%%%%%%%%%%%%
%% MAINMATTER
%%%%%%%%%%%%%%%%%%%%%%%%%%%%%%%%%%%%%%%%%%%%

\begin{quotation}
As far as the laws of mathematics refer to reality, they are not certain, and as far as they are certain, they do not refer to reality. It seems to me that complete clarity [...] first became common property through that trend in mathematics known as ``axiomatics''. The progress achieved by axiomatics consists in its having neatly separated the logical-formal from its objective or intuitive content; according to axiomatics, the logical-formal alone forms the subject matter of mathematics, which is not concerned with the intuitive or other content associated with the logical formal.  \\
\begin{flushright}
Albert Einstein, from ``Geometry and Experience'', a lecture given on January 27, 1921. 
\end{flushright}
\end{quotation}

\vspace{0.7 cm}

\begin{quotation}
To admit singularities does not seem to me the right way. I think that in order to make real progress we must once again find a general principle conforming more truly to nature. \\
\begin{flushright}
Albert Einstein, from a letter to H. Weyl, June 6, 1922. 
\end{flushright}
\end{quotation}

\section{What is ontology?}

Ontology is the more general class of theories about the world. It is more basic than any specific science because it is not concerned with individual entities and their interactions but with very general features of all existents. Ontology deals with questions such as `what is a law of nature?' `what are properties?' `what are events?' `what categories of being there are, if any?' `what is space?' `what is time?' `is the world determined?' and so on. According to Bunge's Dictionary, ontology is ``the branch of philosophy that studies the most pervasive features of reality [...]. Ontology does not study constructs, i.e. ideas in themselves.'' \cite{Bunge1}. 

Ontology, then, is what once was called ``metaphysics''. This name has fallen in disgrace, especially among physicist, since the strong attacks by logical positivists in the first half of the twenty century. `Metaphysics' is sometimes reserved to the great and comprehensive philosophical systems created by Aristotle, Kant, Leibniz, etc. In the past few decades, nonetheless, metaphysics has become respectable once again, among philosophers at least. Sometimes, the name {\em general ontology} is used for metaphysics, and ``ontology'' is restricted to the narrow sense adopted by Quine and others in the late 1940s \cite{Quine1}.

In the latter sense, ontology is understood as the class of entities accepted by a given theory. It is said that if we accept a theory as true, then we are committed to accept the existence of those entities that are {\em truth-makers} of the theory. In Quine's parlance the ontology is the range of the bound variables of a theory. In other words, the ontology is the reference class of the theory. The ontology of biology is the collection of living beings. The ontology of ancient atomism, the collection of all atoms, i.e. indivisible entities. Although all this may seem trivial, when we look at some contemporary and highly formalized theories of physics, the answer to the questions ``what is the associated ontology?'' can be far from obvious. A classic example is Quantum Mechanics. What is Quantum Mechanics about? Particles? Waves? Observers? Instruments? Fields? Probabilities? Minds? Infinite worlds? Quantons? Any physics aficionado knows that the discussions on the particular are never-ending. Without a formal analysis of the foundations of the theory it is very difficult to settle the issue. It is an amazing fact that we can make extraordinary accurate predictions without knowing what we are talking about.

In this article I intend to discuss the ontology of General Relativity. Although this theory is not considered as problematic as Quantum Mechanics, a quick overview of the literature shows that the mere analysis of the natural language used to present the theory is plagued with ambiguities. Is the theory about space-time? Or is it about clocks and rods? Or fields? Or gravitating objects? What about singularities? It is occasionally said that there are theorems in the theory that imply the existence of space-time singularities. Is this correct? In what follows I want to dispel some of the conceptual fog that covers these issues.              

\section{Language, representation, and reality}

The most basic assumption of science is that there is a reality to be known. Without the postulate of the independent existence of a real world the scientific enterprise would be vain. I shall not discuss this primary assumption here. Rather, I want to focus on how we represent the world in our attempts to understand it\footnote{By ``the world'' I understand the totality of existents.}. 

In order to build representations of some features of the world we use conceptual systems called {\em languages}. In ordinary life natural languages such as English, German or Spanish are, or seem to be, enough for most purposes. If we want to penetrate deeper into the structure of reality, however, we need formal languages as those provided by logic and mathematics. 

A formal language is a system of signs with a set of explicit rules to generate valid combinations of symbols (see, for instance, the treatises by Bunge \cite{Bunge2a}, \cite{Bunge2b}  and Martin \cite{Martin}). These rules give instructions ({\em syntactic rules}) about how valid arrangements of symbols (called {\em formulas}) are formed, or relate symbols and formulas with extra-linguistic objects ({\em semantic rules}). The operation of {\em deduction} allows to obtain valid formulas from valid formulas. If a set of formulas is closed under deduction, we call it a {\em theory}. Any {\em interpreted}\footnote{`Interpreted' in this context means `endowed with semantic rules'.} theory, with the help of auxiliary conditions, should produce {\em statements} about states of affairs that occur in the world. If a state of affairs can be used to validate a statement and this statement satisfies truth conditions\footnote{See ref. \cite{Tarski} on formal truth.} in the theory, we say that the theory (actually one of its models) {\em represents} some aspect of the world. 

The use of formal languages in science brings many advantages, to the point that those specific disciplines that do not make intensive use of formal methods are generally underdeveloped. Clarity and precision are gained through formalization. This results in a significant reduction of the vagueness that is inherent to natural languages. The extensive use of formalized languages, besides, enables us to elicit in a systematic way the consequences of our assumptions. The adoption of the special class of formalized languages of mathematics makes possible to introduce quantitative and complex representations of properties and changes that we detect in the world. 

A basic assumption of factual science is that a property can be represented by some mathematical function. Reality is not mathematical, but certainly our more accurate representations of it are. Physical systems, in general, are described by models where properties and processes are represented by mathematical constructs. Models, in this way, are representations of the mechanisms that we assume occur in physical systems. This is so to the extent that {\em to explain} a thing is to unveil the mechanisms that operate in it, i.e. to faithfully represent the manifold of physical processes with a coherent system of mathematical functions and constructs \cite{Bunge3}. 

Sometimes, in highly elaborated theories, however, formalization can reach such a degree of complexity that the semantics of the language might be difficult to elucidate. This results in problems of interpretation and is a source of confusion. 

\section{The structure of physical theories}

When a theory becomes too complex, axiomatization provides a way of clarification, making all assumptions and definitions explicit. Since every theory is expressed in some language, it can be arranged as a system of statements and definitions. More exactly a theory $T$ is a set of statements $s$ closed under the operation of logical implication ($\rightarrow$): any $s\in T$ is either an axiom (basic statement) or a consequence of the set of axioms $A$ ($T=\{s : A \rightarrow s\}$).

The axioms can be grouped in three subsets: mathematical axioms, physical axioms, and semantic axioms \cite{Bunge0}. The first group states the basic formal structure of the theory. For instance, in a field theory mathematical axioms introduce a set with a given topology and some additional structure (e.g. metric) that allows to define a mathematical space. The semantical axioms relate the mathematical structures with physical entities. These axioms equip the theory with an interpretation. This interpretation and the meanings of the different terms are propagated from the axioms (with the help of adequate definitions that allow for economy of language) to the theorems or derived statements. A formal theory of semantics must be applied to determine the flow of meaning from the basis $A$ to any $s\in T$.

Finally, the physical axioms express relations among constructs that represent physical entities. These relations are usually presented in the form of differential or integro-differential equations that represent physical laws, i.e. objective patterns in the world\footnote{Some physicists claim that ``laws are expressed by functionals'' but this is utterly wrong. Functionals are constructs that can be used to obtain laws through some general principles, such as the Principle of Minimal Action. Functionals do not express laws by themselves. For example, a Lagrangian does not represent a law for a field, the law is obtained through the Lagrange equations.}. If a given function represents a physical property of a material system, then physical axioms (equations) provide constraints to the state space accessible to the system. The physical axioms are the core of the theory.       

In addition to its formal, semantic, and physical structure, any theory has some background knowledge that is assumed. This background is a set of formal theories (e.g. topology and differential geometry), physical theories (e.g. electromagnetism) and purely ontological theories (e.g. space-time, in the case of General Relativity). We turn now to the latter theory, fundamental for the understanding of gravitation.  

\section{What is space-time?}

General Relativity is said sometimes to be a ``a theory of space and time''. We read, for instance, in the classic textbook by Misner, Thorne, and Wheeler \cite{Gravitation}: ``Space acts on matter, telling it how to move. In turn, matter reacts back on space, telling it how to curve"'. The concept of space-time, however, is presupposed by all classical field theories. The concept was introduced by Minkowski \cite{Mink}, and belongs to ontology, not to physics \cite{Romero1}. A formal construction of space-time can be obtained starting from an ontological basis of either things \cite{Bunge77}, \cite{Bergliaffa3} or events \cite{Romero2}. In what follows I provide a simple outline of an ontological theory of space-time. Let us begin defining space-time.\\

{\em Space-time is the emergent of the ontological composition of all events}. \\

Events can be considered as primitives or can be derived from things as changes in their properties if things are taken as ontologically prior. Both representations are equivalent since things can be construed as bundles of events \cite{Romero2}. Since composition is not a formal operation but an ontological one\footnote{For instance, a human body is composed by cells, but is not just a mere collection of cells since it has emergent properties and specific functions far more complex than those of the individual components.}, space-time is neither a concept nor an abstraction, but an emergent entity. As any entity, space-time can be represented by a concept. The usual representation of space-time is given by a 4-dimensional real manifold $E$ equipped with a metric field $g_{ab}$:

$$  
{\rm ST}\hat{=}\left\langle E, g_{ab}\right\rangle.
$$

I insist: space-time {is not} a manifold (i.e. a mathematical construct) but the ``totality'' of all events. A specific model of space-time requires the specification of the source of the metric field. This is done through another field, called the ``energy-momentum'' tensor field $T_{ab}$. Hence, a model of space-time is:

$$  
M_{\rm ST}=\left\langle E, g_{ab}, T_{ab}\right\rangle.
$$

The relation between both tensor fields is given by field equations. The metric field specifies the geometry of space-time. The energy-momentum field represents the potential of change (i.e. event generation) in space-time. 

We can summarize all this in the following axioms.\\

${\rm P1 - Syntactic}.$ The set $E$ is a $C^{\infty}$ differentiable, 4-dimensional, real pseudo-Riemannian manifold.\\

${\rm P2 - Syntactic}. $ The metric structure of $E$ is given by a tensor field of rank 2, $g_{ab}$, in such a way that the differential distance $ds$ between two events is: $$ds^{2}=g_{ab} dx^{a} dx^{b}.$$

${\rm P3 - Syntactic}.$ The tangent space of $E$ at any point is Minkowskian, i.e. its metric is given by a symmetric tensor $\eta_{ab}$ of rank 2 and trace $-2$.\\

${\rm P4 - Syntactic}.$ The metric of $E$ is determined by a rank 2 tensor field $T_{ab}$ through the following field equations:

\begin{equation}
G_{ab}-g_{ab}\Lambda=\kappa T_{ab}, \label{Eq-Einstein} 
\end{equation}
where $G_{ab}$ is the Einstein tensor (a function of the second derivatives of the metric). Both $\Lambda$ and $\kappa$ are constants.\\

${\rm P5 - Semantic}.$ The elements of $E$ represent physical events.\\

${\rm P6 - Semantic}.$ Space-time is represented by an ordered pair $\left\langle E, \; g_{ab}\right\rangle$: $${\rm ST}\hat{=}\left\langle E, g_{ab}\right\rangle.$$

${\rm P7 - Semantic}.$ There is a non-geometrical field represented by a 2-rank tensor field $T_{ab}$ on the manifold E.\\ 

${\rm P8 - Semantic}.$ A specific model of space-time is given by: $$M_{{\rm ST}}=\left\langle E, g_{ab}, T_{ab}\right\rangle.$$\\

Notice that so far no mention has been done of the gravitational field. The sketched theory is purely ontological, and hence, cannot be identified yet with General Relativity.    

\section{The gravitational field and space-time}

The ontological theory of space-time is not a theory of the gravitational field. It is a theory presupposed by all theories of physics. Any representation of dynamical processes takes place `in' space-time, i.e. it is parametrized in terms of space-time variables. In order to describe gravitation through a space-time theory, additional semantic rules are necessary. These rules will link space-time concepts with gravitational concepts. In words of Hans Reichenbach ``It is not the theory of gravitation that becomes geometry, but it is geometry that becomes an expression of the gravitational field'' \cite{Reichenbach}. Some mathematical (geometric) objects of the space-time formalism represent some aspect of the gravitational field. But, what object represents the gravitational field?  

The Einstein tensor is:
\begin{equation}
	G_{ab}\equiv R_{ab}-\frac{1}{2}R g_{ab},
\end{equation}
where $R_{ab}$ is the Ricci tensor formed from second derivatives of the metric and $R\equiv g^{ab}R_{ab}$ is the Ricci scalar. The geodetic equations for a test particle free in the gravitational field is:
\begin{equation}
	\frac{d^{2}x^{a}}{d\lambda^{2}}+ \Gamma^{a}_{bc}\frac{dx^{b}}{d\lambda}\frac{dx^{c}}{d\lambda},
\end{equation}
with $\lambda$ an affine parameter and $\Gamma^{a}_{bc}$ the affine connection, given by:
\begin{equation}
\Gamma^{a}_{bc}=\frac{1}{2}g^{ad}(\partial_{b}g_{cd}+\partial_{c}g_{bd}-\partial_{d}g_{bc}).	
\end{equation}
  
The affine connection is not a tensor, but can be used to build a tensor that is directly associated with the curvature of space-time: the Riemann tensor. The form of the Riemann tensor for an affine-connected manifold can be obtained through a coordinate transformation 
${x^{a}\rightarrow {\bar{x}^{a}}}$ that makes the affine connection  vanish everywhere, i.e.

\begin{equation}
	\bar{\Gamma}^{a}_{bc}(\bar{x})=0, \;\;\; \forall\; \bar{x},\; a,\;b,\; c.
\end{equation}

\vspace{0.2cm}

\noindent The coordinate system ${\bar{x}^{a}}$ exists if

\begin{equation}
\Gamma^{a}_{bd, c}-\Gamma^{a}_{bc, d} + \Gamma^{a}_{ec}\,\Gamma^{e}_{bd} - \Gamma^{a}_{de}\,\Gamma^{e}_{bc}=0
\label{R}
\end{equation}

\noindent for the affine connection $\Gamma^{a}_{bc}({x})$. The left hand side of Eq. (\ref{R}) is the Riemann tensor:
\begin{equation}
	R^{a}_{bcd}=\Gamma^{a}_{bd, c}-\Gamma^{a}_{bc, d} + \Gamma^{a}_{ec}\,\Gamma^{e}_{bd} - \Gamma^{a}_{de}\,\Gamma^{e}_{bc}.
\end{equation}
 
When $R^{a}_{bcd}=0$ the metric is flat, since its derivatives are zero. If \mbox{$K=R^{a}_{bcd}R^{bcd}_{a}>0$} the metric has a positive curvature. Sometimes is said that the Riemann tensor represents the gravitational field, since it only vanishes in the absence of fields. On the contrary, the affine connection can be set locally to zero by a transformation of coordinates. This fact, however, only reflects the equivalence principle: the gravitational field can be suppressed in any locally free falling system. Wording differently, the tangent space to the manifold that represents space-time is always Minkowskian. To determine the mathematical object of the theory that represents the gravitational field we have to consider the weak field limit of Eqs. (\ref{Eq-Einstein}). When this is done we find that that the gravitational potential is identified by the metric coefficient $g_{00}\approx \eta_{00} + h_{00}$ and the coupling constant $\kappa$ is $-8\pi G/c^{4}$. If {\em the metric represents the gravitational potential}, then {\em the affine connection represents the strength of the field itself}. This is similar to what happens in electrodynamics, where the 4-vector $A^{\mu}$ represents the electromagnetic potential and the tensor field $F^{\mu\nu}=\partial_{\mu}A_{\nu}-\partial_{\nu}A_{\mu}$ represents the strength electromagnetic field. {\em The Riemann tensor, by other hand, being formed by derivatives of the affine connection, represents the rate of change, both in space and time, of the strength of the gravitational field} \cite{Lehmkuhl}.

The source of the gravitational field in Eqs. (\ref{Eq-Einstein}), the tensor field $T_{ab}$,  represents also physical properties of material things. It represents the energy and momentum of all non-gravitational systems. 

The class of reference of General Relativity is then formed by all gravitational fields and all physical systems that generate such fields. This means that, since all material objects have energy and momentum, the reference class of the theory is maximal: it includes everything. This, of course, does not imply that the theory provides a correct description of any physical system in the universe, as we shall see. Moreover, we shall argue in what follows that the theory is necessarily incomplete.        

\section{Are there singularities?}
 
Singularities are features of some solutions of Eqs. (\ref{Eq-Einstein}). Hence, they appear in some models $M_{\rm ST}$ of space-time. Several authors have inferred from this that singularities should exist in the world, i.e. that they are part of the ontology of General Relativity. I disagree.  

A space-time model is said to be {\sl singular} if its manifold $E$ is {\sl incomplete}. A manifold is incomplete if it contains at least one {\sl inextendible} curve. A curve $\gamma:[0,a)\longrightarrow E$ is inextendible if there is no point $p$ in $E$ such that $\gamma(s)\longrightarrow p$ as $a\longrightarrow s$, i.e. $\gamma$ has no endpoint in $E$. It is said that a given space-time model $\left\langle E, \;g_{ab}\right\rangle$ has an {\sl extension} if there is an isometric embedding $\theta: M\longrightarrow E^{\prime}$, where $\left\langle E^{\prime}, g_{ab}^{\prime}\right\rangle$ is another space-time model and $\theta$ is an application onto a proper subset of $E^{\prime}$. {\em Essential} singular space-time models do not admit regular extensions; they essentially contains at least one curve $\gamma$ that is inextendible in the sense given above. The singular character of the solution cannot be avoided just finding an adequate coordinate system.  

Singular space-times are said to contain singularities, but this is an abuse of language: singularities are not `things' in space-time, but a pathological feature of some solutions of the fundamental equations of the theory.  

An essential or true singularity should not be interpreted as a representation of a physical object of infinite density, infinite pressure, etc. Since the singularity does not belong to the manifold that represents space-time in General Relativity, it simply cannot be described or represented in the framework of such a theory. General Relativity is incomplete in the sense that it cannot provide a full description of the gravitational behavior of any physical system \cite{Kundt}. True singularities are not within the range of values of the bound variables of the theory: they do not belong to the ontology of a world that can be described with 4-dimensional differential manifolds. 

Several singularity theorems can be proved from pure geometrical properties of the space-time model \cite{Clarke1993}, \cite{Hawking-73-Cambridge}. The most important of these theorems is due to Hawking \& Penrose \cite{H-P}:\\

{\bf Theorem.} Let $\left\langle E,\;g_{ab}\right\rangle$ be a time-oriented space-time satisfying the following conditions:
\begin{enumerate}
	\item $R_{ab}V^{a}V^{b}\geq 0$ for any non space-like $V^{a}$.
	\item Time-like and null generic conditions are fulfilled.
	\item There are no closed time-like curves.
	\item At least one of the following conditions holds
	
\begin{itemize}
	\item a.  There exists a compact\footnote{A space is said to be compact if whenever one takes an infinite number of "steps" in the space, eventually one must get arbitrarily close to some other point of the space. For a more formal definition, see below. 
%Thus, whereas disks and spheres are compact, infinite lines and planes are not, nor is a disk or a sphere with a missing point. In the case of an infinite line or plane, one can set off making equal steps in any direction without approaching any point, so that neither space is compact. In the case of a disk or sphere with a missing point, one can move toward the missing point without approaching any point within the space. More formally,  a topological space is compact if, whenever a collection of open sets covers the space, some sub-collection consisting only of finitely many open sets also covers the space. A topological space is called compact if each of its open covers has a finite sub-cover. Otherwise it is called non-compact. Compactness, when defined in this manner, often allows one to take information that is known locally -- in a neighborhood of each point of the space -- and to extend it to information that holds globally throughout the space.
} achronal set\footnote{A set of points in a space-time with no two points of the set having time-like separation. } without edge.
	\item b. There exists a trapped surface.
	\item c. There is a $p\in E$ such that the expansion of the future (or past) directed null geodesics through $p$ becomes negative along each of the geodesics.  
\end{itemize}
\end{enumerate}
 
Then, $\left\langle E,\;g_{ab}\right\rangle$ contains at least one incomplete time-like or null geodesic. \\

If the theorem has to be applied to the physical world, the hypothesis must be supported by empirical evidence. Condition 1 will be satisfied if the energy-momentum $T_{ab}$ satisfies the so-called {\em strong energy condition}: $T_{ab}V^{a}V^{b}\geq -(1/2)T^{a}_{a}$, for any time-like vector $V^{a}$. If the energy-momentum is diagonal and corresponds to an ideal fluid, the strong energy condition can be written as $\rho+3 p\geq 0$ and $\rho + p\geq 0$, with $\rho$ the energy density and $p$ the pressure. Condition 2 requires that any time-like or null geodesic experiences a tidal force at some point in its history. Condition 4a requires that, at least at one time, the universe is closed and the compact slice that corresponds to such a time is not intersected more than once by a future directed time-like curve. The trapped surfaces mentioned in 4b refer to horizons due to gravitational collapse.  Condition 4c requires that the universe is collapsing in the past or the future. 
 
The theorem is purely geometric, no physical law is invoked. Theorems of this type are a consequence of the gravitational focusing of congruences. A congruence is a family of curves such that exactly one, and only one, time-like geodesic trajectory passes through each point $p\in E$. If the curves are smooth, a congruence defines a smooth time-like vector field on the space-time model. If $V^{a}$ is the time-like tangent vector to the congruence, we can write the {\em spatial part} of the metric tensor as:
\begin{equation}
	h_{ab}=g_{ab}+V_{a}V_{b}.
\end{equation}
 
For a given congruence of time-like geodesic we can define the {\em expansion}, {\em shear}, and {\em vorticity} tensors as:

\begin{eqnarray}
\theta_{ab}&=&V_{(i;l)}h^{i}_{a}h^{l}_{b}, \\	
\sigma_{ab}&=&\theta_{ab}-\frac{1}{3}h_{ab}\theta,  \\
\omega_{ab}&=&h^{i}_{a}h^{l}_{b}V_{[i;l]}.  
\end{eqnarray}
Here, the {\em volume expansion} $\theta$ is defined as:

\begin{equation}
	\theta=\theta_{ab}h^{ab}=\nabla_{a}V^{a}=V^{a}_{\;\;\; ;a}.
\end{equation}

The rate of change of the volume expansion as the time-like geodesic curves in the congruence are moved along is given by the Raychaudhuri \cite{Ray} equation:

	\[\frac{d\theta}{d\tau}=-R_{ab}V^{a}V^{b}-\frac{1}{3}\theta^{2}-\sigma_{ab}\sigma^{ab}+\omega_{ab}\omega^{ab},
\]

or 

\begin{equation}
\frac{d\theta}{d\tau}=-R_{ab}V^{a}V^{b}-\frac{1}{3}\theta^{2}-2\sigma^{2}+2\omega^{2}.	
\end{equation}

We can use now the Einstein field equations (without $\Lambda$) to relate the congruence with the space-time curvature:

\begin{equation}
	R_{ab}V^{a}V^{b}= \kappa \left[T_{ab}V^{a}V^{b}+\frac{1}{2} T\right]. \label{cond-R}
\end{equation}

The term $T_{ab}V^{a}V^{b}$ represents the energy density measured by a tie-like observer with unit tangent four-velocity $V^{a}$. The weak energy condition then states that:
\begin{equation}
	T_{ab}V^{a}V^{b}\geq 0. 
\end{equation}
 
The strong condition is:
\begin{equation}
	T_{ab}V^{a}V^{b}+\frac{1}{2} T\geq 0. 
\end{equation}
Notice that this condition implies, according to Eq. (\ref{cond-R}),
\begin{equation}
	R_{ab}V^{a}V^{b}\geq 0.
\end{equation}

We see, then, that the conditions of the Hawking-Penrose theorem imply that the focusing of the congruence yields:

\begin{equation}
	\frac{d\theta}{d\tau}\leq -\frac{\theta^{2}}{3}, \label{teor-sing}
\end{equation}
where we have used that both the shear and the rotation vanishes. Equation (\ref{teor-sing}) indicates that the volume expansion of the congruence must necessarily decrease along the time-like geodesic. Integrating, we get:
\begin{equation}
	\frac{1}{\theta}\geq \frac{1}{\theta_{0}}+ \frac{\tau}{3},
\end{equation}
where $\theta_{0}$ is the initial value of the expansion. Then, $\theta\rightarrow -\infty$ in a finite proper time $\tau\leq 3/\left|\theta_{0}\right|$. This means that once a convergence occurs in a congruence of time-like geodesics, a caustic (singularity) must develop in the space-time model. The non space-like geodesics are in such a case inextendible.  

Singularity theorems are not theorems that imply the physical existence, under some conditions, of space-time singularities. Material existence cannot be formally implied. Existence theorems imply that under certain assumptions there are functions that satisfy a given equation, or that some concepts can be formed in accordance with some explicit syntactic rules. Theorems of this kind state the possibilities of some formal system or language. The conclusion of the theorems, although not obvious in many occasions, is always a necessary consequence of the assumptions made. 

In the case of singularity theorems of classical field theories like General Relativity, what is implied is that under some assumptions the solutions of the equations of the theory are defective beyond repair. The correct interpretation of these theorems is that they point out the {\em incompleteness} of the theory: there are statements that cannot be made within the theory. In this sense (and only in this sense), the theorems are like G\"odel's famous theorems of mathematical logic\footnote{G\"odel's incompleteness theorems are two theorems of mathematical logic that establish inherent limitations of all but the most trivial axiomatic systems capable of doing arithmetic. The first theorem states that any effectively generated theory capable of expressing elementary arithmetic cannot be both consistent and complete \cite{Godel}.}.  

To interpret the singularity theorems as theorems about the existence of certain space-time models is wrong. Using elementary second order logic is trivial to show that there cannot be non-predicable objects (singularities) in the theory \cite{Romero3}. If there were a non-predicable object in the theory,

\begin{equation}
	\left(\exists x\right)_{E} \; \left(\forall P\right) \sim Px, \label{P}
\end{equation}
where the quantification over properties in unrestricted. The existential quantification $\left(\exists x\right)_{E}$, on the other hand, means

$$\left(\exists x\right)_{E} \equiv \left(\exists x\right) \wedge \left( x\in E\right).$$ 

Let us call $P_{1}$ the property `$x\in E$'. Then, formula (\ref{P}) reads:
\begin{equation}
\left(\exists x\right)	\left(\forall P\right) (\sim Px \; \wedge P_{1}x ), \label{P1}
\end{equation}
which is a contradiction, i.e. it is false for any value of $x$. 

We conclude that there are no singularities nor singular space-times. There is just a theory with a restricted range of applicability. 

\section{The continuum approximation and its breakdown}

The existence of singular solutions in a background-independent\footnote{A background-independent theory lacks any fixed, non-dynamical space-time structure.} theory such as General Relativity is a consequence of some contradiction at the level of the axiomatic base of the theory. Such contradiction manifests itself as essential divergences in the solutions. I maintain that the contradiction arises from the continuum approximation adopted to model the gravitational field through a metric field on a manifold and the attractive characteristic of gravity at small scales, that leads to accumulation points and the consequent apparition of divergences. This strongly suggest that the continuum approximation fails to describe space-time at small scales. The field description requires infinite degrees of freedom, that are not available at certain scales. A short distance cut-off in the degrees of freedom should prevent the emergence of singular space-times. Discreteness must be added as an independent assumption to the axioms of the theory when dealing with very short range phenomena. However, this cannot be done in the framework of classic General Relativity. Rather, a discrete theory should be developed from which General Relativity (and the usual notions of space and time) can emerge as some kind of averages, in similar way to the emergence of thermodynamics from mechanical statistics. This implies a major ontological shift. Before discussing the ontological implications, I find convenient to review the continuum approximations of General Relativity.      

We have represented the ontological composition of all events by a set $E$ and we have mentioned that this set has the structure of a differentiable manifold. Let us review the implications. 

\subsection{Topological spaces}

Let $E$ be any set and $T=\{E_{\alpha}\}$ a collection, finite or infinite, of subsets of $E$. Then $(E,\; T)$ is a {\em topological space} iff:
\begin{enumerate}

\item $E\in T$.
\item $\emptyset \in T$.
\item Any finite or infinite sub-collection $\{E_{1}, E_{2}, ..., E_{n}\}$ of the $E_{\alpha}$ is such that $\bigcup^{n}_{1} E_{i}\in T$. 
\item Any {\em finite} sub-collection $\{E_{1}, E_{2}, ..., E_{n}\}$ of the $E_{\alpha}$ is such that $\bigcap^{n}_{1} E_{i}\in T$.   
\end{enumerate}

The set $E$ is called a topological space and the $E_{\alpha}$ are called {\em open sets}. The assignation of $T$ to $E$ is said to `give' a topology to $E$. 

A function $f$ mapping from the topological space $E$ onto the topological space $E^{\ast}$ is continuous if the inverse image of an open set in $E^{\ast}$ is an open set in $E$.

If a set $E$ has two topologies $T^{1}=\{E_{\alpha}\}$ and $T^{2}=\{E^{\ast}_{\alpha}\}$ such that $T^{1}\supset T^{2}$, we say that $T^{1}$ is {\em stronger} than $T^{2}$.

%\subsection*{Neighborhoods}

Given a topology $T$ on $E$, then $N$ is a {\em neighborhood} of a point $e\in E$ if $N\subset E$ and there is some $E_{\alpha} \subset N$ such that $e\in E_{\alpha}$. Notice that it is not necessary for $N$ to be an open set. However, all open sets $E_{\alpha}$ which contain $e$ are neighborhoods of $e$ since they are contained in themselves. Thus, neighborhoods are more general than open sets. 

%\subsection*{Closed sets}

Let again $T$ be a topology on $E$. Then any $U\subset E$ is {\em closed} if the complement of $U$ in $E$ ($\bar{U}=E-U$) is an open set. Since $\bar{\bar{U}}=U$ then a set is open when its complement is closed. The sets $E$ and $\emptyset$ are open and closed regardless the topology $T$.  

%\subsection*{Closure of a set}

Given a set $U$, there will be in general many closed sets that contain $U$. Let ${F_{\alpha}}$ be the family of closed set that contain $U$. The {\em closure} of $U$ is $\tilde{U}=\bigcap_{\alpha} F_{\alpha}$. The closure is the smallest closed set that contains $U$. Notice that $\tilde{\tilde{U}}=\tilde{U}$.  

%\subsection*{Boundary and interior}

The {\em interior} $U^{0}$ of a set $U$ is the union of all open sets $O_{\alpha}$ of $U$: $U_{0}=\bigcup_{\alpha} O_{\alpha}$. The interior of $U$ is the largest open set of $U$. 

The {\em boundary} $b(U)$ of a set $U$ is the complement of the interior of $U$ in the closure of $U$:
$b(U)=\tilde{U}-U^{0}$. Closed sets always contain their boundaries:

$$ U\cap b(U) = \emptyset \Longleftrightarrow U \; {\rm is\; open.}$$

$$b(U)\subset U \Longleftrightarrow U \; {\rm is\; closed.}   $$

%Notice that the sets $(a,\;b)$, $[a,\;b)$, $(a,\;b]$, and $[a,\;b]$ all have the same boundary: %$b={a,\;b}$.

%\subsection*{Compactness}

A subset $U$ is said {\em dense} in $X$ if $\tilde U=X$.

Given a family of sets $\{F_{\alpha}\}=F$, $F$ is a {\em cover} of $U$ if $U\subset\bigcup_{\alpha} F_{\alpha}$. If $(\forall F_{\alpha})_{F} (F_{\alpha} {\rm \;is\; an\; open\; set})$ then the cover is called an {\em open cover}. 

A set $U$ is {\em compact} if for {\em every} open covering $\{F_{\alpha}\}$ with $U\subset \bigcup_{\alpha} F_{\alpha}$ there {\em always} exist a {\em finite} sub-covering $\{F_{1},...,F_{n}\}$ of $U$ such that $\bigcup^{n}_{1}F_{\alpha}\subset U$.

As an illustration consider the $n$-dimensional real space $\Re^{n}$. A subset $X$ of $\Re^{n}$ is compact iff it is closed and bounded. This means that $X$ must have finite area and volume in $n$-dimensions.   

%\subsection*{Connectedness}

A set $E$ is {\em connected} if it cannot be written as: $E=E_{1}\cup E_{2}$ where $E_{1}$ and $E_{2}$ are both open sets and $E_{1}\cap E_{2}=\emptyset$.

%\subsection*{Homeomorphisms and topological invariants}

Let $T_{1}$ and $T_{2}$ be two topological spaces. An {\em homeomorphism} is a map $f$ from $T_{1}$ to $T_{2}$: 
$$ f: T_{1}\rightarrow T_{2}$$
such that $f$ is continuous and its inverse map $f^{-1}$ is also continuous. If there is a third topological space $T_{3}$ such that $T_{1}$ is homeomorphic to $T_{2}$ and $T_{2}$ is homeomorphic to $T_{3}$, then $T_{1}$ is homeomorphic to $T_{3}$. An homeomorphism defines an equivalence class, that of all spaces that homeomorphic to a given topological space. If the homeomorphism $f$ and its inverse $f^{-1}$ are infinitely differentiable ($C^{\infty}$), the $f$ is called a {\em diffeomorphism}. All diffeomorphisms are homeomorphisms, but the converse is not always the case.  

A {\em topological invariant} is a construct that does not change under homeomorphisms. They are characteristics of the equivalence class of the homeomorphism. An example of an invariant is the dimension $n$ of $\Re^{n}$. 

Homeomorphisms generate equivalence classes whose members are topological spaces. Instead, {\em homotopies} generate classes whose members are continuous maps. More specifically, let $f_{1}$ and $f_{2}$ be two continuous maps between the topological spaces $T_{1}$ and $T_{2}$:
$$f_{1}: T_{1}\rightarrow T_{2},  $$ 
$$f_{2}: T_{1}\rightarrow T_{2}.  $$
Then $f_{1}$ is said to be homotopic to $f_{2}$ if $f_{1}$ can be deformed into $f_{2}$. Formally:
$$F: T_{1}\times[0,\;1]\rightarrow T_{2}, \;\;\; F\;{\rm continuous}$$ and
$$F(x,\; 0)=T_{1}(x), $$
$$F(x,\; 1)=T_{2}(x). $$ 

This means that as the real variable $t$ changes continuously from 0 to 1 in the interval $[0,\;1]$ the map $f_{1}$ is deformed continuously in the map $f_{2}$. Homotopy is an equivalence relation that divides the space of continuous maps from $T_{1}$ to $T_{2}$ into equivalent classes. These homotopy equivalent classes are topological invariants of the pair of spaces $T_{1}$ and $T_{2}$.

Homotopy can be used to classify topological spaces. If we identify one of the topological spaces with the n-dimensional sphere $S^{n}$, then the space of continuous maps from $S^{n}$ to $T$, $C(S^{n}, \; T)$, can be divided into equivalence classes according to the topological space $T$. The equivalent classes of $C(S^{n}, \; T)$ have a group structure and form the homotopy group $\Pi_{n}(T)$.

It is rather straightforward to show that both compactness and connectedness are topological invariants.

\subsection{Manifolds: definition and properties}

A set $E$ is a differentiable manifold if:
\begin{enumerate}
\item $E$ is a topological space.
\item $E$ is equipped with a family of pairs $\{(E_{\alpha}, \; \phi_{\alpha})\}$.
\item The $E_{\alpha}$'s are a family of open sets that cover $E$: $E=\bigcup_{\alpha} E_{\alpha}$. The $\phi_{\alpha}$'s are homeomorphisms from $E_{\alpha}$ to open subsets $O_{\alpha}$ of $\Re^{n}$: $\phi_{\alpha}: E_{\alpha}\rightarrow O_{\alpha}$.
\item Given $E_{\alpha}$ and $E_{\beta}$ such that $E_{\alpha}\cap E_{\beta}\neq \emptyset$, the map $\phi_{\beta} \circ \phi^{-1}_{\alpha}$ from the subset $\phi_{\alpha}(E_{\alpha}\cap E_{\beta})$ of $\Re^{n}$ to the subset $\phi_{\beta}(E_{\alpha}\cap E_{\beta})$ of $\Re^{n}$ is infinitely differentiable ($C^{\infty}$). 
\end{enumerate}

The family $\{(E_{\alpha},\;\phi_{\alpha})\}$ is called an {\em atlas}. The individual members of the atlas are {\em charts}. In informal language we can say that $E$ is a space that can be covered by patches
$E_{\alpha}$ which are assigned coordinates in $\Re^{n}$ by $\phi_{\alpha}$. Within each of these patches $E$ looks like a subset of the Euclidean space $\Re^{n}$. $E$ is not necessarily globally Euclidean or pseudo-Euclidean. If two patches overlap, then in $E_{\alpha}\cap E_{\beta}$ there are two assignations of coordinates, which can be transformed smoothly into each other. The dimension of the manifold $E$ is the dimension $n$ of the space $\Re^{n}$.

A manifold $E$ is said to be Hausdorff if for any two distinct elements $x\in E$ and $y\in E$, there exist $O_{x}\subset E$ and $O_{y}\subset E$ such that $O_{x} \cap O_{y}=\emptyset$.

A given topological space $E$ is said to be {\em metric} if there is a map $d:E\times E \rightarrow \Re$ such that, for any $x,\;y,$ and $z$ in $E$:
\begin{enumerate}
\item $d(x,\; y)\geq 0$.
\item $d(x,\; y)= 0$ iff $x=y$.
\item If $z\in E$, then $d(x,\; y)+d(y,\; z)=d(x,\; z)$.  
\end{enumerate} 

An important property of manifolds is their {\em orientability}. Given a manifold $E$ whose atlas is $\{(E_{\alpha}, \;\phi_{\alpha})\}$, $E$ is orientable if ${\rm det}(\phi_{\beta}\circ \phi^{-1}_{\alpha})>0$ for all $E_{\alpha}$ and $E_{\beta}$ such that $E_{\alpha}\cap E_{\beta}\neq\emptyset$. The manifold is orientable if one can define a preferred direction unambiguously. 

The manifolds adopted in General Relativity to represent space-time have a pseudo-Riemannian metric and are compact. A very important property is that every metric space is compact if and only if every subset has at least one accumulation point. \\

{\bf Definition.} Let $E$ be a topological space and $A$ a subset of $E$. A point $a\in A$ is called an {\em accumulation point} of $A$ if each neighborhood of $a$ contains infinitely many points of $A$.\\

For compact Hausdorff spaces, every infinite subset $A$ of $E$ has at least one accumulation point in $E$. 
 
If we want to represent events at very small scale, compactness must be abandoned. The reason is that any accumulation point implies an infinite energy density, since events are represented by points of the manifold, events have finite (but not arbitrarily small) energy, and energy is an additive property. In other words, space-time must be discrete at the smallest scale.   

\section{Quantum Gravity: a theory about...what?}

As far as we can decompose a given event $e\in E$ into more basic events, in such a way that $E$ can be approximated by a compact uncountable (non-denumerable) metric space, the continuum representation for the totality of events will work. But if there are atomic\footnote{I use the word `atomic' in the original Greek sense of $\acute{\alpha}\tau o\mu o\varsigma$, ``uncut'', ``individual'', ``not decomposable''.} events, there will be a sub-space of $E$ that is countable (or denumerable if it is infinite) and ontologically basic. There is, in such a case, a discrete substratum underlying the continuum manifold. Since the quantum of action is given by the Planck constant, it is a reasonable hypothesis to assume that the atomic events occur at the Planck scale, $l_{\rm P}=\sqrt{\hbar G/c^{3}}$. If there are atomic events, the way they associate would give rise to composed events (i.e. processes), and then to the continuum space-time, which would be a large-scale emergent property, absent at the more basic ontological level. This is similar to, for instance, considering the mind as a collection of complex processes of the brain, emerging from arrays of `mindless' neurons.

If this view is correct, then quantum gravity is a theory about relations among basic events and the ontological emergence of space-time and gravity. Quantum gravity would be a theory so basic that it might well be considered as ontological rather than physical.   

The discrete nature of the space-time ontological substratum can be formed by atomic events. It has been suggested \cite{Sorkin1} that such events and their relations should be represented by a partially ordered set ({\em poset}). It can be proved that the dimension, topology, differential structure, and metric of the manifold where a poset is embedded is determined by the poset structure \cite{Malament}. If the order relation is interpreted as a causal relation, the posets are called {\em causal sets} (or {\em causets}). We do not need to make this distinction at this level.

A poset is a set $P$ with a partial order binary relation $\preceq$ that is:
\begin{itemize} 
\item Reflexive: For all $x \in P$, $ x \preceq x$.
\item Antisymmetric: For all $x,\; y\; \in P$, $x \preceq y \preceq x$ implies $x = y$.
\item Transitive: For all $x,\; y,\; z \in P$, $x \preceq y \preceq z$ implies $x \preceq z$.
\item Locally finite: For all $x,\; z \in P$, {\bf card} $(\{y \in C | x \preceq y \preceq z\}) < \infty$.
\end{itemize}

Here {\bf card} $(A)$ denotes the {\em cardinality} of the set $A$. Notice that $x \prec y$  if $x \preceq y$ and $x \neq y$.

The causal relation of a Lorentzian manifold (without closed causal curves) satisfies the first three conditions. It is the local finiteness condition that introduces space-time discreteness. A given poset can be embedded into a Lorentzian manifold. An embedding is a map taking elements of the poset into points in the manifold such that the order relation of the poset matches the causal ordering of the manifold. A further criterion is needed however before the embedding is suitable. If, on average, the number of causal set elements mapped into a region of the manifold is proportional to the volume of the region, the embedding is said to be faithful. The poset is then called {\em manifold-like}

A conjecture (called {\it hauptvermutung}) must be made to ensure that the same poset cannot be faithfully embedded into two space-times that are not similar on large scales. 

Alternatively, a poset can be generated by {\em sprinkling} points (events) from a Lorentzian manifold. By sprinkling points in proportion to the volume of the space-time regions and using the causal order relations in the manifold to induce order relations between the sprinkled points, a poset can be produced that (by construction) can be faithfully embedded into the manifold. 

To maintain Lorentz invariance this sprinkling of points must be done randomly using a Poisson process. Thus, the probability of sprinkling $n$ points (events) into a region of volume $V$ is:
\begin{equation}
P(n) = \frac{(\rho V)^n e^{-\rho V}}{n!},
\end{equation}
where $\rho$ is the density of the sprinkling.

A {\em link} in a poset is a pair of elements $x,\; y \in P$ such that $x \prec y$ but with no $z \in 
P$ such that $x \prec z \prec y$. In other words, $x$ and $y$ represent directly linked events. 

A {\em chain} is a sequence of elements $x_0,\; x_1,\ldots,x_n$ such that $x_i \prec x_{i+1}$ for $i=0,\ldots,n-1$. The length of a chain is $n$, the number of relations used. A chain represents a process.

A geodesic between two poset elements can then be introduced as follows: a geodesic between two elements $x,\; y \in P$ is a chain consisting only of links such that $x_0 = x$ and $x_n = y$. The length of the chain, $n$, is maximal over all chains from $x$ to $y$. In general there will be more than one geodesic between two elements. The length of a geodesic should be directly proportional to the proper time along a timelike geodesic joining the two space-time points if the embedding is faithful.

\begin{figure}
\scalebox{0.40}{\includegraphics{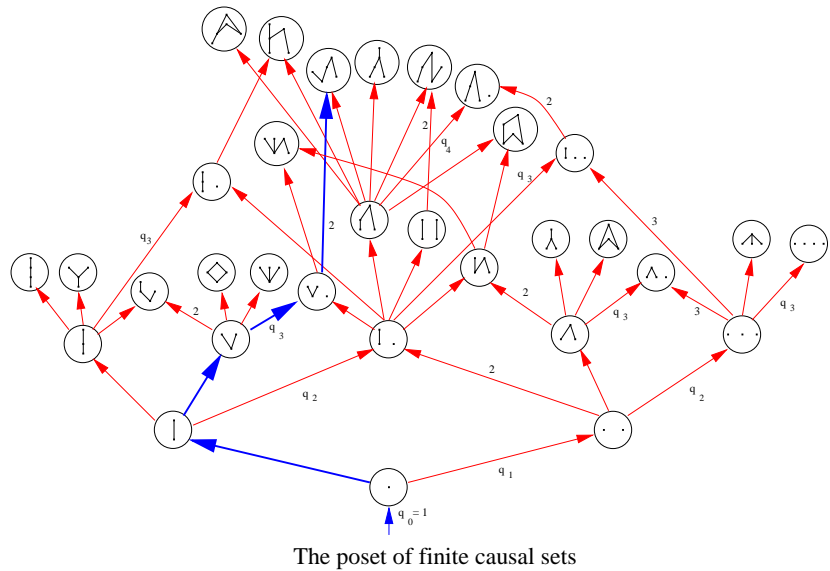}} \label{poscau}
\caption{Poset generation by classical sequential growth (from Ref. \cite{Walden})}.
\end{figure}

A major challenge is to recover a realistic space-time structure starting from a numerable poset. This problem is sometimes called ``dynamics of causets''. A step in the direction of solving the problem  is a classical model in which elements are added according to probabilities. This model is known as classical sequential growth (CSG) dynamics \cite{Sorkin2}. The classical sequential growth model is a way to generate causal sets by adding new elements one after another (see Figure \ref{poscau}). Rules for how new elements are added are specified and, depending on the parameters in the model, different causal sets result. The direction of growing gives rise to time, which does not exist at the fundamental poset event level.

Another challenge is to account for the remaining referents of General Relativity, namely, gravitating objects. I suggest that physical objects can be understood as clusters of processes, and hence can emerge as inhomogeneities in the growing pattern of events. This conjecture is supported by the observation that whatever exists seems to have energy, and energy is just the capability to change. The most populous the bundle of events, the larger the associated energy. In this view, energy emerges as well, just as a measure of the density of basic events. Objects, physical things, would be nothing else than clusters of events.

The ontology of Quantum Gravity, and of the world, under this perspective, would be a maelstrom of basic events; the things, people, galaxies of the universe, arise as a patterned poset in that storm.

\section{Conclusions}

General Relativity is a theory about the gravitational field and the physical systems that interact with it, i.e. all physical systems. General Relativity is {\em not} a theory about space-time. Space-time is an ontological emergent property of the system formed by all existents, whatever they are. Because of the unique universality of gravity, models of space-time can be used to represent the gravitational field in General Relativity. More precisely, the affine connection of space-time represents the strength of the field, and the metric represents the gravitational potential. 

Space-time emerges from events and processes. Processes are nothing else than chains of events. The most basic events, at the Planck scale, can be supposed to form a numerable set with a partial order, a poset. These posets represent the basic substratum of reality. The space-time continuum emerges from the basic discrete events. From a formal point of view, the manifold model of General Relativity should be recovered from an adequate large number limit of a discrete set theory endowed with a partial order structure. Time, as well as the spatial dimensions, should emerge in the process as well, along with the topological and metric properties we attribute to space-time. Bridging this breach between the discrete and the continuum, between the basic and the complex, between atomic events and our daily reality, is perhaps the greatest scientific enterprise of our time.

\section*{Appendix: Axiomatic General Relativity}

In this Appendix we provide an axiomatic formulation of General Relativity. Previous axiomatizations have been presented by Bunge \cite{Bunge0} and Covarrubias \cite{Covarrubias}.

The formal background of the theory is formed by first order logic, set theory, topology, and differential geometry. The physical background consists of general system theory and Special Relativity. The ontological background is space-time theory, as sketched above. 

The primitive base of basic concepts that appear in the axiomatic base is given by:
$$\left\langle M^{n}, \; \left\{{\bf g}\right\},\; \Sigma, \; \bar{\Sigma}, \;\left\{ {\bf T} \right\}, \;K, \;\Lambda,\;\kappa \right\rangle. $$
Here, $M^{n}$ is a $n$-dimensional real differential manifold, $\; \left\{{\bf g}\right\}$ is a family of pseudo-Riemannian metric fields on $M^{n}$, $\Sigma$ is the collection of gravitational fields, $\bar{\Sigma}$ is the collections of physical entities other than gravitational fields, $\left\{ {\bf T} \right\}$ is a family of 2-covariant tensor fields functionals of ${\bf g}$ and possibly of state variables, $K$ is the collection of all possible reference systems, $\Lambda$ is a constant, and $\kappa$ is a negative real dimensional number.

The full meaning of these symbols is established through the following axiom system.

\begin{itemize}

\item {\bf A.1} $\Sigma$ is a nonempty collection such that every $\sigma\in \Sigma$ designates a gravitational field.

\item {\bf A.2} $\bar{\Sigma}$ is a nonempty collection such that every $\bar{\sigma}\in \bar{\Sigma}$ designates macroscopic physical system other than a gravitational field. This collection includes a null (fictional) individual that represents the absence of material systems other than gravitational fields. 

\item {\bf A.3} The pseudo-Riemannian $C^{\infty}$ metric manifold $(M^{n}, \; {\bf g})$, with $n=4$ represents space-time.

\item {\bf A.4} $K\neq \emptyset$ $\wedge$ $K\subset \bar{\Sigma}$.

\item {\bf A.5} Every $k\in K$ is a possible physical reference frame. For every such frame a specific coordinate assignation can be defined.

\item {\bf A.6} Each ${\bf T} \in \left\{{\bf T}\right\}$ is a symmetric 2-covariant tensor field, functional of ${\bf g}$ and possibly of state variables over $M^{4}$.  

\item {\bf A.7} $\left(\forall \bar{\sigma}\right)_{\bar{\Sigma}} \left(\exists {\bf T}\right)_{\left\{{\bf T}\right\}} ({\bf T}$ represents the energy, momentum, and stress of $\bar{\sigma})$.

\item {\bf A.8} $\left(\forall \sigma\right)_{\Sigma} \left(\exists {\bf g}\right)_{\bf g} \left(\exists \bar{\sigma}\right)_{\bar{\Sigma}} \left({\bf G}- \Lambda {\bf g}=\kappa {\bf T} \right)$, where ${\bf G}$ is the Einstein tensor formed by second derivatives of ${\bf g}$ and ${\bf T}$ is the energy-momentum tensor field corresponding to $\bar{\sigma}$. 

\item {\bf A.9} $\Lambda$ is a constant that represents the energy density of space-time in the absence of non-gravitational fields. The constant $\kappa$ represents the coupling of the gravitational field with the non-gravitational systems.

\item {\bf A.10} $k=-8\pi G c^{-4}$, with $G$ the gravitational constant and $c$ the speed of light in vacuum. 

\end{itemize}

From these 10 axioms is possible to deduce all the standard theorems of General Relativity \cite{Covarrubias}. Some clarifications are relevant. 

The macroscopic physical systems are partitioned into gravitational and non-gravitational. All systems generate gravitational fields; hence the field equations are non-linear. Since gravitational fields can exist in absence of sources, the null individual (a fiction) is included in $\bar{\Sigma}$. 

Physical systems are grouped in collections, not sets, since they can appear and disappear (i.e. the number of elements in a collection can change, contrary to what happens with set members). The tensor fields that represent physical entities are then grouped in collections as well. 

The existence of free gravitational fields implies that the dimensionality of space-time should be 4 o larger. This can be established as a theorem. We have fixed $n=4$ since we are axiomatizing standard General Relativity.

From the axioms is clear that General Relativity is about the members of both $\Sigma$ and $\bar{\Sigma}$. The theory is not about space-time. The whole ontological theory of space-time, as outlined above, is presupposed by General Relativity. The axioms that link mathematical constructs such as ${\bf g}$ and ${\bf T}$ with referents of the theory are of semantic nature. Once the interpretation is fixed in this way, axiom {\bf A.8} expresses a law of nature. Such a law can be derived from meta-laws, such as the Action Principle, given the right Lagrangian.    

Finally, I remark that General Relativity makes no assumption on the nature of the elements of $\bar{\Sigma}$ and how to represent their energy and momentum, beyond the weak hypothesis that this representation should came in the form of a second rank tensor field of $C^{1}$ class (this can be proved as theorem). The exact form of this field is left to our inferences from the world.

%% The option [height=...] scales the picture to the given height,
%% without it it would be printed at its nominal size
%%%%%%%%%%%%%%%%%%%%%%%%%%%%%%%%%%%%%%%%%%%%

%%%%%%%%%%%%%%%%%%%%%%%
%% SAMPLE TABLE
%%
%% Shows the use of \tablehead and \tablenote
%% macros
%%%%%%%%%%%%%%%%%%%%%%%%%%%%%%%%%%%%%%%%%%%%
\begin{theacknowledgments}
I am very grateful to Prof. Mario Novello for his kind invitation to participate of this stimulating school. I thank Professors Mario Bunge, Santiago Perez Bergliaffa, and Felipe Tovar Falciano, as well as Lic. Daniela P\'erez  for constructive criticisms and comments. I was supported by research grant PIP 0078 from CONICET.

\end{theacknowledgments}




%\bibliography{sample}

%  \typeout{******************************************}
%  \typeout{** Please run "bibtex \jobname" to optain}
%  \typeout{** the bibliography and then re-run LaTeX}
%  \typeout{** twice to fix the references!}
%  \typeout{******************************************}

\end{document}